\documentclass[12pt]{article}
\pdfoutput=1
\usepackage{float}
\usepackage{authblk}
\usepackage{amssymb,amsmath}
\usepackage{graphicx}
\usepackage{caption}
\usepackage{subcaption}
\usepackage{cite}
\usepackage{color}
\usepackage{setspace}
\usepackage[colorlinks=true
,urlcolor=blue
,citecolor=blue
,linkcolor=blue
,pagecolor=blue
,linktocpage=true
,pdfproducer=medialab
]{hyperref}
\usepackage{cleveref}
\usepackage[a4paper,width=15cm,height=25cm]{geometry}
\makeatletter \renewcommand{\@dotsep}{10000} \makeatother

\title{Yukawa Unification with light supersymmetric particles 
consistent with LHC constraints}

\date{}

\author{Mureed Hussain\thanks{Email: mureed.hussain@sns.nust.edu.pk}~}
\author{Rizwan Khalid\thanks{Email: rizwan@sns.nust.edu.pk}}

\vspace{0.25cm}

\affil{Department of Physics, School of Natural Sciences, National University of Sciences \& Technology, H-12, Islamabad, Pakistan}

\begin{document}
\maketitle
\begin{abstract}
We investigate supersymmetric models with left-right symmetry based on the group $SU(4)_{c} \times SU(2)_{L} \times SU(2)_{R}$ ($4$-$2$-$2$) with negative sign of bilinear Higgs potential parameter $\mu$ in the context of the latest experimental results. In the backdrop of experimental results from the Large Hadron Collider, we investigate the possibility of Yukawa unification in $4$-$2$-$2$ and find out the same is still not ruled out. Furthermore, this scenario also provides a satisfactory dark matter candidate. The current experimental bounds on sparticle masses, mass bounds on Higgs particle, updated phenomenological constraints from the rare decays of B meson and the anomalous magnetic moment of muon with the requirement of a Yukawa unified theory having $10 \%$ or better third family Yukawa unification are utilized to bound the parametric space of these models.    
\end{abstract}

\section{Introduction}
The successful gauge couplings unification in the Minimal Supersymmetric Standard Model (MSSM) at the scale $M_{GUT}\simeq 2 \times 10^{16}$ GeV provides a hint for the existence of a supersymmetric (SUSY) grand unified theory (GUT). In SUSY GUT models based on, for example, the gauge group SO(10)\cite{Fritzsch:1974nn, GellMann:1976pg, Hagiwara:2002fs, Auto:2003ys} some additional unification can be provided like all matter of a single generation is unified into the 16 dimensional spinor multiplet in addition to predicting the existence of a right handed neutrino which naturally leads to neutrino masses via the Type-I see-saw mechanism \cite{GellMann:1980vs}. In the simplest SO(10) SUSY GUT models, the 10-dimensional Higgs multiplet contains the Standard Model (SM) Higgs doublets and also predicts the Yukawa coupling unification for the third generation as the superpotential of such models contains the term $W=Y$ $\bf{16}$ $\bf{10}$ $\bf{16}$ which results in $Y_{t}=Y_{b}=Y_{\tau}=Y_{\nu_\tau}=Y$. The viability of the third family (t-b-$\tau$) Yukawa coupling unification in supersymmetric models has been explored in many papers(see for example \cite{Auto:2003ys, Dermisek:2018hxq}). The t-b-$\tau$ Yukawa unification, for superpotential Higgs mass parameter $\mu>0$, to be less then $1 \%$ is possible for the universal scalar mass parameter $m_{0} \sim 8-20$ TeV and for the small values of the universal gaugino mass $M_{1/2}\lesssim 400$ GeV in order to also accommodate neutralino dark matter via gluino coannihilation scenario \cite{Auto:2003ys}. These models also require a GUT scale mass splitting between the Higgs scalars with $m^{2}_{H_{u}}< m^{2}_{H_{d}}$. If $\mu <0$ then t-b-$\tau$ Yukawa unification can occur for $m_{0}$ and $M_{1/2}$ $\gtrsim$ $1-2$ TeV. With all these intriguing factors, it is difficult to address the dark matter relic abundance in SO(10) models as it requires $m_{0}$ to be lighter than 3 TeV for $\mu>0$ and $M_{1/2}$ to be  heavier than 2.5 TeV for $\mu<0$\cite{Auto:2003ys} and the only possibility in this restricted scenario is that of light Higgs resonance\cite{Baer:2008jn}. 

 Precision studies of rare decays of B mesons proves to be a popular as well as a very much useful mechanism to put constraints on any physics beyond the SM~ \cite{Hurth:2006mm}. A new era for these studies has begun since the Large Hadron Collider beauty (LHCb) experiment started providing data for different observables for these decays. This data has been providing increasingly stringent constraints on the MSSM~\cite{Arbey:2012bp, Hussain:2017fbp}. Constraints from rare B decays like~ $B_{s} \rightarrow \mu^{+} \mu^{-}$, $B_{u} \rightarrow \tau \nu_{\tau}$, $b \rightarrow s \gamma$ are routinely used in studying bounds on the parametric spaces of different SUSY models (see, for example, \cite{Arbey:2012bp, Hussain:2017fbp} and references therein). For the rare semi-leptonic decay  ($B_{d} \rightarrow K^{*} \mu^{+} \mu^{-}$), LHCb has also provided a set of quantum chromodynamics (QCD) form factor independent observables~\cite{Aaij:2013iag}. So far no $5\sigma$ deviation from the SM has been shown in B Physics constraints so they suggest lower bounds models incorporating new physics.

The anomalous magnetic moment of the muon, $a_{\mu}=(g-2)_{\mu}/2$, has been measured with significant precision by the Muon $g$-$2$ Collaboration~\cite{BNL}. The SM prediction for the anomalous magnetic moment of the muon,  \cite{Gohn:2016ezs}, has a $3.5 \sigma$ discrepancy with the experimental results \cite{BNL}. An important observable has been defined as the difference between experimental value of $g_{\mu} -2$ and its theoretical value calculated in the SM, given as $ \Delta a_{\mu} = a_{\mu}^{\text{exp}} - a_{\mu}^{\text{SM}}$~(hereafter referred to as the $g_\mu-2$ anomaly). The experimental value for this observable is
\begin{equation}
\Delta a_{\mu}= (28.6 \pm 8.0) \times 10^{-10}.
\end{equation}
The requirement of accommodating the experimental value of $g_\mu$ in a new physics scenario, such as low scale supersymmetry~\cite{Ajaib:2015yma}, yields an upper and a lower bound for such a theory. Several studies has been subjected to probe that whether the $g_\mu-2$ anomaly can be resolved, provided that we also simultaneously satisfy the constraints from direct searches for SUSY particles and indirect searches through B Physics~\cite{Wang:2018vrr, Hussain:2017fbp}.

The discovery of the Higgs particle of mass $\sim125$ GeV~\cite{:2012gk,:2012gu} at the LHC has important effects on low scale supersymmetry. In order to have such a heavy Higgs particle in the MSSM a couple of alternatives may work. We can either require a large, ${\cal O} (\mathrm{few}-10)$ TeV stop squark or introduce a relatively large soft supersymmetry breaking (SSB) trilinear $A_t$-term \cite{Heinemeyer:2011aa}. This, of course, translates into 
restricting and constraining the parameters of any model.

In this paper we study the question of Yukawa coupling unification within the context of a maximal subgroup of $SO(10)$ based on the $SU(4)\times SU(2)\times SU(2)$ gauge symmetry also known as the Pati-Salam or 4-2-2 model. We investigate Yukawa coupling unification as motivated by the 4-2-2 model in light of LHC bounds and investigate whether we can have neutralino dark matter in such models in addition to satisfying the constraints on $\Delta a_\mu$. In Section~\ref{4-2-2} we give a brief introduction to the 4-2-2 model that we use. In Section~\ref{scanProc}, we describe details of the $4$-$2$-$2$ model parameters in the MSSM language along with our scanning procedure and details of the constraints we impose. We present our results in Section~\ref{results} and in Section~\ref{conclusion} we conclude our discussion. 
\section{4-2-2 Model}
The gauge symmetry $SU(4)_{c} \times SU(2)_{L} \times SU(2)_{R}$ ($4$-$2$-$2$) is a maximal subgroup of $SO(10)$ and exhibits many essential features of its covering group. As a standalone symmetry group, 4-2-2 implements electric charge quantization in units of $e/6$. It also assigns quark and lepton families in bi-fundamental representations and, in addition, predicts the existence of right handed neutrinos.
It has been well known that the left-right symmetric models can produce the SM gauge structure \cite{Melfo:2003xi},
\begin{equation}
SU(2)_{L} \times SU(2)_{R} \times U(1)_{B-L} \to SU(2)_{L} \times U(1)_{Y}
\end{equation}  
These models also provide a natural explanation for baryon and lepton number conservation. Charge quantization can be explained if $U(1)_{B-L}$ is formed by a $SU(4)_{c}$ group using the fact that  $SU(3)\times U(1)$ $\subset$ $SU(4)$ . This set of models are called Pati-Salam models and have the following breaking pattern,
\begin{eqnarray} 	
SU(4)_{c} \times SU(2)_{L} \times SU(2)_{R}
&\to&  SU(3)_{c} \times SU(2)_{L} \times SU(2)_{R} \times U(1)_{B-L} \nonumber \\ 
&\to& SU(3)_{c} \times SU(2)_{L} \times U(1)_{Y}. 
\end{eqnarray}\label{4-2-2}
In the 4-2-2 model the $16$-plet of $SO(10)$ matter fields and their corresponding conjugate fields consist of $\psi(4,2,1)$ and $\psi_{c}(\bar{4},2,1)$. In this model the third family Yukawa couplings $\psi_{c}H \psi$, where $H(1,2,2)$ denotes the bi-doublet, yields the following relation valid at $M_{GUT}$ \cite{Gogoladze:2009ug},
\begin{equation}
Y_{t}=Y_{b}=Y_{\tau}=Y_{\nu_{\tau}}.
\end{equation} 
Providing a discrete left-right (LR) symmetry \cite{Pati:1974yy} (more precisely C-parity\cite{Kibble:1982ae}) to the 4-2-2 imposes the gauge coupling unification condition ($g_{L}=g_{R}$) at the GUT scale ($M_{GUT}$) which results in the reduction of the number of independent gauge couplings from three to two. Due to this C-parity, the soft-symmetry-breaking (SSB) terms which are induced at $M_{GUT}$ through gravity mediated supersymmetry breaking are equal in magnitude for squarks and sleptons of the three families. Gaugino masses at $M_{GUT}$ due to this C-parity, associated with $SU(2)_{L}$ and $SU(2)_{R}$ are expected to be the same ($M_{2}\equiv M^{L}_{2} =M^{R}_{2}$) while the gaugino masses associated with color symmetry $SU(4)_{c}$ can be different. So we have two independent parameters ($M_{2}$ and $M_{3}$) in the gaugino sector for the supersymmetric 4-2-2 model. The relation between the three MSSM gaugino SSB masses is given by
\begin{equation}
M_{1}=\frac{3}{5}M_2 + \frac{2}{5}M_3.
\end{equation}
It has been shown in a previous study \cite{Gogoladze:2010fu} that the t-b-$\tau$ Yukawa unification requires relatively large threshold corrections to $Y_{b}$ which come mainly from the gluino and chargino loops~\cite{Pierce:1996zz}. For $\mu > 0$, the required contributions need a large $m_{0}$ and $A_{t}$ where $A_{t}$ is the top trilinear coupling. For large $m_{0}$ the gluino and chargino contribution scales as $M_3/m^{2}_{0}$ and $A_{t}/m^{2}_{0}$, respectively. On the other hand, for $\mu < 0$, Yukawa unification can be achieved for significantly low $m_{0}$ and also for a wider range of $A_{t}$ because the gluino term, contributing to threshold correction to $Y_{b}$, gets the required negative sign \cite{Gogoladze:2010fu}.

The supersymmetric contributions to the $g_{\mu}-2$ are proportional to $\mu M_{2}$ so if we take $\mu< 0$, the SUSY corrections to $g_{\mu}-2$ get the incorrect sign(assuming $M_{2}$ $>$ $0$) which is the reason for the choice of $\mu >0$ in most of the studies of physics beyond SM. In the $4$-$2$-$2$ model, as we can have different mass parameters for gauginos so we can have a negative sign of $\mu$-parameter if we take negative $M_{2}$.

The 4-2-2 model has been extensively investigated since its proposal~\cite{Pati:1974yy}. In \cite{Gogoladze:2009ug}, it has been shown that in the 4-2-2 model, Yukawa unification is consistent with the neutralino dark matter abundance and collider bounds (except $g_{\mu}-2$). The bounds on MSSM sfermions mass parameter $``m_{0}"$ is $8$~$TeV$. Gluino and bino-wino coannihilation regions which are consistent with the WMAP dark matter constraints are shown in \cite{Gogoladze:2009bn}. In \cite{Gogoladze:2010fu} it has been shown that if the sign of MSSM bi-linear Higgs mass parameter $\mu$ is taken to be negative along with negative $M_{2}$ and positive $M_{3}$ then there are regions of parametric space of 4-2-2 model which are consistent with collider bounds including $g_{\mu}-2$ anomaly constraint and with 10 $\%$ or better Yukawa unification along with a sub TeV sparticle spectrum. Constraints from the branching ratios of rare B decays $B_{s} \to \mu^+ \mu^{-}$, $b \to s \gamma$ and $B_{u} \to \tau \nu_{\tau}$ were also applied in \cite{Gogoladze:2009ug, Gogoladze:2009bn, Gogoladze:2010fu }. However, 
no analysis was done vis a vis the angular and 
other observables in the $B\to K^*$ sector. Further, the bounds applied in these studies have now long been updated by the LHC. In a more recent article \cite{Poh:2017xvg} a statistical analysis of $SO(10)$ breaking via Pati-Salam model based on the 4-2-2 gauge symmetry together with a family symmetry is done and the consistency of theoretical bounds to the experimental results are discussed requiring $m_{0} \gtrsim 15$ TeV.

In this article we apply constraints on the 4-2-2 parametric space from the current collider bounds provided by the ATLAS and CMS experiments. We also apply the $g_{\mu}-2$ anomaly constraint along with the constraints form the rare decays of B mesons. We have included the constraints from the angular observables and the zero crossing of the forward-backward asymmetry of rare B decay $B_{d} \to K^{*} \mu^{+} \mu^{-}$. We use data provided by the LHCb experiment for B physics observables. We also analyse the sparticles mass bounds in context of Yukawa unification and also provide the expected mass ranges for them. We have chosen $\mu<0$ in order to get as light a spectrum as possible and can achieve Yukawa unification consistent with constraints for $m_0 \lesssim 5$ TeV as shown in Section~\ref{results}.

\section{Scanning procedure and constraints}\label{scanProc}
We do random scans over the parameter space of $4$-$2$-$2$ model by using the SOFTSUSY-3.5.2 package \cite{SOFTSUSY}. This package calculates the sparticle spectrum in the CP conserving MSSM with full flavor mixing structure and solves the renormalization group equations with boundary conditions the SSB terms specified at $M_{GUT}$. We use the fermion mass data and weak scale gauge couplings as boundary conditions at energy scale equal to $M_Z$ (the $Z$ boson mass). The $4$-$2$-$2$ parameter space that we have scanned is,
\begin{equation}\label{pars}
\begin{array}{rcl}
0 ~\le &m_{0}& \le ~5 ~\text{TeV}, \\
-5 ~\le &M_{2}& \le ~ 0 ~\text{TeV}, \\
0 ~\le &M_{3}& \le ~ 5 ~\text{TeV},\\
-3~\le &A_{0}/m_{0}& \le ~ 3,  \\
0~\le &\tan\beta & \le ~60,\\
0~\le &m_{H_{u}}& \le ~5 ~\text{TeV},\\
0~\le &m_{H_{d}}& \le ~5 ~\text{TeV},\\
&\mu&<~ 0,
\end{array}
\end{equation}
where $m_{0}$ is the sfermion mass parameter, $M_{2}$ and $M_{3}$ are the gaugino mass parameters as described in the previous section, $A_{0}$ is the universal trilinear coupling parameter, $\tan \beta$ is the ratio of the vacuum expectation values of the two MSSM Higgs doublets, and $\mu$ is the supersymmetric bilinear Higgs parameter whose square is, of course, fixed by the requirement of radiative electroweak symmetry breaking. 

After the generation of the sparticle spectrum by SOFTSUSY (written out in the SUSY LesHouches Accord (SLHA)~\cite{Skands:2003cj} format), we use the SUPERISO package~\cite{Mahmoudi:2008tp} to calculate different B physics observables. The theoretical value of $g_\mu -2$ is also calculated by SUPERISO package. 

We then apply the constraints from the lower mass bounds on sparticle masses which are provided by the ATLAS and the CMS experiments~\cite{Aad:2012fqa}. These mass bounds are,
\begin{equation}\label{gluino}
\left.
\begin{array}{rcl}
m_{\tilde{g}} &>& 1900 \hspace{0.2cm} \text{GeV} , \\
m_{\tilde{q}}  &>&  1600 \hspace{0.2cm}  \text{GeV},\\
m_{\tilde{\chi}^{0}_{1}} & >& 46 \hspace{0.2cm}  \text{GeV},\\
m_{\tilde{\chi}^{0}_{2}} & >& 670 \hspace{0.2cm}  \text{GeV} 
\hspace{0.2cm}(m_{\tilde{\chi}^{0}_{1}} < 200 \hspace{0.2cm}  \text{GeV}),\\
m_{\tilde{\chi}^{0}_{2}} & >& 116 \hspace{0.2cm}  \text{GeV},\\
m_{\tilde{\chi}^{\pm}_{1}} & >& 103 \hspace{0.2cm}  \text{GeV} 
\hspace{0.2cm}(m_{\tilde{\nu}} > 300 \hspace{0.2cm}\text{GeV}),\\
m_{\tilde{\chi}^{\pm}_{1}} & >& 94 \hspace{0.2cm}  \text{GeV},
\end{array}
\right\}
\end{equation}
where $m_{\tilde{g}}$ is mass of gluino(the supersymmetric partner of SM gluon), $m_{\tilde{q}}$ is mass of squarks (lightest of the squarks of first two families) , $m_{\tilde{\chi}^{0}_{i}}$ are the neutralino masses and $m_{\tilde{\chi}^{\pm}_{i}}$ are the chargino masses. $m_{\tilde{\nu}}$ is the mass of the sneutrinos.   

We impose the following mass bound on the SM-like Higgs,  
\begin{equation}\label{higgs}
123.0 \hspace{0.1cm} \text{GeV} \hspace{0.2cm} \le m_{h^{0}} \le \hspace{0.2cm} 127.0 \hspace{0.1cm} \text{GeV}.
\end{equation}

We next apply constraints from different observables calculated for rare B decays. The branching ratios of rare $B$ decays $B_{s} \rightarrow \mu^{+} \mu^{-}$, $b \rightarrow s \gamma$ and $B_{u} \rightarrow \tau \nu_{\tau}$ have been used to constrain the MSSM~\cite{Ellis:2007fu}. In this study, we also include these decays for constraining the parametric space of $4$-$2$-$2$ model. The experimental values for these branching ratios are given as 
\begin{equation}\label{oldB}
\left.
\begin{array}{rcl}
BR(B_s \rightarrow \mu^{+}\mu^{-})& = & (2.9 \pm 0.7)\times 10^{-9}, \\
BR(b \rightarrow s \gamma) & = & (3.43 \pm 0.22 )\times 10^{-4},  \\
\frac{BR(B_{u} \rightarrow \tau \nu_{\tau})_{\text{MSSM}}}{BR(B_{u} \rightarrow \tau \nu_{\tau})_{\text{SM}}} &=& 1.13 \pm 0.43.
\end{array}
\right\}
\end{equation}
For the rare semi-leptonic decay $B_{d} \rightarrow K^{*} \mu^{+} \mu^{-}$, we can measure the differential branching ratio in terms of square of the dilepton invariant mass ($q^{2}$)\cite{Chatrchyan:2013cda}. Some additional observables for this decay channel like the longitudinal polarization fraction ($F_{L}$)\footnote{It is the ratio of the decay rate of $B_{d} \rightarrow K^{*} \mu^{+} \mu^{-}$ when $K^{*}$ is longitudinally polarized to the total decay rate of $B_{d} \rightarrow K^{*} \mu^{+} \mu^{-}$.} of $K^{*}$ and forward-backward asymmetry ($\mathcal{A}_{FB}$)\footnote{It is measured in terms of number of events in which final state leptons are moving in forward or backward direction in the rest frame of decaying $B$ meson.} are also being measured by the LHCb experiment\cite{Chatrchyan:2013cda}. The zero crossing of the $\mathcal{A}_{FB}$ is also an important observable as it puts constraints on a variety of models~\cite{Mahmoudi:2014iia}. For this decay mode, a plethora of experimental observables are provided by the differential decay distribution of the four-body final state ($B \to K^{*}  \ell^{+}\ell^{-}$ (where $K^{*} \to K \pi$)) ~\cite{Straub:2015ica} as the associated differential decay rate can be written in terms of four kinematic observables ($q^2$ the invariant mass square of lepton, and three angles $\theta_{l}$, $\theta_{K}$ and $\phi$), with coefficients $J_{i}$\footnote{For details about kinematics of this decay we refer to~\cite{Chatrchyan:2013cda, Matias:2012xw}}. These $J_{i}$ depend on the transversity amplitudes and their explicit form is given in \cite{Matias:2012xw}. The $J_{i}(q^2)$ when integrated in different $q^2$ bins form some important observables which are now measured at the LHCb experiment. These observables also contain the hadronic uncertainties due to QCD effects which can be minimized by taking the appropriate ratios of these $J_{i}'s$. These ratios are known as the optimized observables for this decay which are then measured in different $q^{2}$ bins. The optimized observables are denoted by $P_i$. In~ \cite{Descotes-Genon:2013wba}, the authors have done a detailed discussion on choosing these observables in light of results from the LHCb. This decay, due to its rich kinematics, has proven to be very important as it is used to constrain new physics scenarios~\cite{Hurth:2016fbr}.

In our study, we calculate these observables for each point of the $4$-$2$-$2$ parametric space and then compare them to experimental results by calculating $\chi^2$ which is given as:
\begin{eqnarray}\label{chiSquare}
\chi^2 = &&\sum\limits_{\text{bins}}\bigg[\sum\limits_{{i,j}\in (B \to K^{*} \mu ^+ \mu^- obs.)} 
(O^{\text{exp}}_{i}- O^{\text{th}}_{i})
(\sigma^{(\text{bin})})^{-1}(O^{\text{exp}}_{j}- O^{\text{th}}_{j})\bigg] \nonumber\\ 
&&+ \sum\limits_{k \in (\text{other B physics obs.})}\frac{(O^{\text{exp}}_{k}- O^{\text{th}}_{k})}{(\sigma^{\text{exp}}_{k}- \sigma^{\text{th}}_{k})} + \sum\limits_{l \in (\Delta a_{\mu}, q^{2}_{0})} \frac{(O^{\text{exp}}_{l}- O^{\text{th}}_{l})}{(\sigma^{\text{exp}}_{l})^2},
\end{eqnarray}
where $O^{exp}$ and $O^{th}$ are the experimental and theoretical values of the corresponding observable, respectively whereas $\sigma$ represents its standard deviation. A particular $p\text{-value}$ is calculated for each data point which is then used to measure the confidence level (CL) by $(1-p) \times 100$. First term in Eq.\eqref{chiSquare} is the contribution to the $\chi^2$ from the optimized observables for $B_{d} \to K^{*} \mu^{+} \mu^{-}$ decay while the second term gives the contribution from the other B physics observables given in Eq.\eqref{oldB}. Third term in Eq.\eqref{chiSquare} gives the contribution of constraints from $g_{\mu}-2$ anomaly and from zero crossing of the forward-backward asymmetry of $B_{d} \to K^{*} \mu^{+} \mu^{-}$ decay. We have done the $\chi^2$ analysis using by techniques by authors of~\cite{Hurth:2016fbr} and refer the same for more details. 
\newpage
\section{Results}
\label{results}
We use a parameter $R_{tb\tau}$ to quantify t-b-$\tau$ Yukawa unification as
\begin{equation}
R_{tb\tau}=\frac{\text{max}(Y_{t},Y_{b}, Y_{\tau})}{\text{min}(Y_{t},Y_{b}, Y_{\tau})},
\end{equation}
and so $R_{tb\tau}=1$ means perfect Yukawa coupling unification whereas, $R_{tb\tau}=1.05$ can be said to correspond to Yukawa unification within $5\%$, for example.
\begin{figure}[H]
	\captionsetup{width=0.90\textwidth}
	\begin{subfigure}{.5\textwidth}
		\includegraphics[width=8cm,height=6cm]{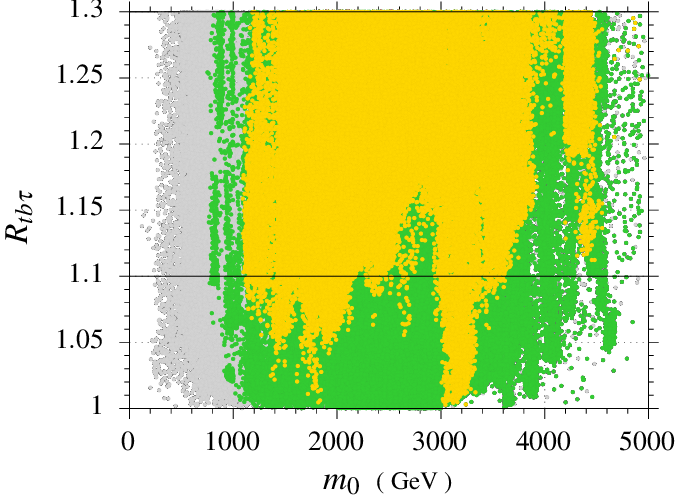}
		\label{fig:sfig11}
	\end{subfigure}
	\begin{subfigure}{.5\textwidth}
		\includegraphics[width=8cm,height=6cm]{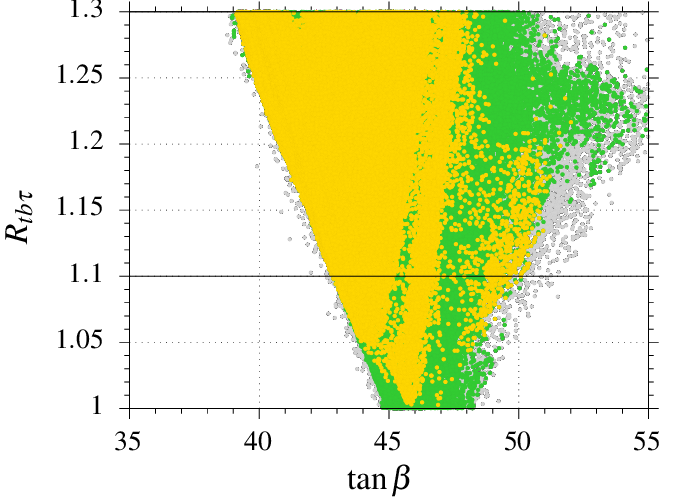}
		\label{sfig12}
	\end{subfigure}
	\begin{subfigure}{.5\textwidth}
		\includegraphics[width=8cm,height=6cm]{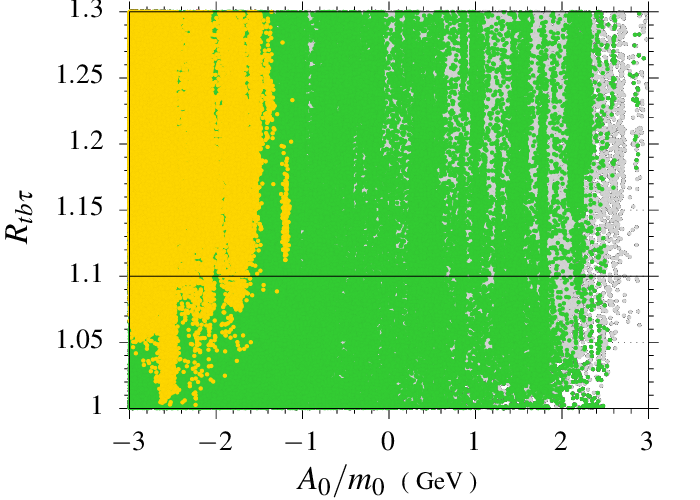}
		\label{fig:sfig13}
	\end{subfigure}
	\begin{subfigure}{.5\textwidth}
		\includegraphics[width=8cm,height=6cm]{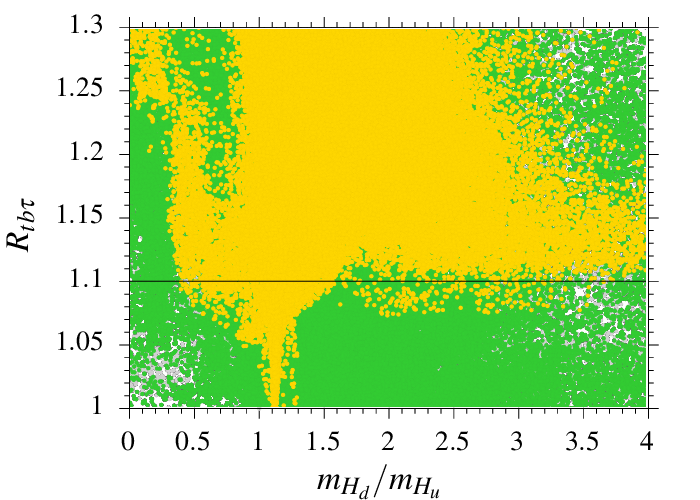}
		\label{fig:sfig14}
	\end{subfigure}
	\caption{Gray points are in accordance with the REWSB and neutralino LSP conditions.
		Green points form a subset of gray points and satisfy mass bounds on sparticles and SM-like Higgs particle. 
		Yellow points are a subset of green points and satisfy the constraints from the $g_{\mu}-2$ anomaly
		and B physics observables within 95$ \% $ confidence level region.}
	\label{fig:fig1}
\end{figure}
In Fig. (\ref{fig:fig1}), we show our plots in the $R_{tb\tau}$-$m_{0}$, $R_{tb\tau}$-$\tan\beta$ $R_{tb\tau}$-${A_{0}/m_{0}}$ and $R_{tb\tau}$-$m_{H_{d}}/m_{H_{u}}$ planes. Gray points satisfy the radiative electroweak symmetry breaking (REWSB) condition. Gray points also satisfy the condition that the lightest supersymmetric particle (LSP) is a neutralino so it could be a viable dark matter candidate. Green points form a subset of gray points and they additionally satisfy the neutralino, chargino, gluino and squark mass bounds provided by the ATLAS and the CMS experiments as given in Eq.\eqref{gluino}. Green points also satisfy the Higgs mass bounds given in Eq.\eqref{higgs}. Yellow points form a subset of green points and lie in the $95 \%$ confidence level range of $\chi^2$. The $\chi^2$ is calculated for all the B physics constraints including branching ratios of $b \to s \gamma$, $B_{s} \to \mu^+ \mu^-$, $B_{u} \to \tau \nu_{\tau}$. The calculated $\chi^2$ also includes contribution from the zero crossing of the forward-backward asymmetry and from the angular observables ${P'}_{i}s$ for semileptonic decay $B \to K^* \mu^+ \mu^-$ decay. The contribution from $g_{\mu}-2$ anomaly constraint is also included while calculating this $\chi^2$.

In $R_{tb\tau}$ - $m_{0}$ plane we can see that the sparticles and B physics along-with $g_{\mu}-2$ constraints put very stringent lower bounds on $m_{0}$ even if we do not consider Yukawa unification. The sparticle mass bound constraints put a lower bound of $\sim 750$ GeV while B physics and $g_{\mu}-2$ constraints further push this limit to $1050$ GeV.

In $R_{tb\tau}$ - $\tan \beta$ plot we can see that for perfect Yukawa unification $\tan \beta$ should be in the range of $\sim 45-50$.
We can also see that the constraints from sparticles mass bounds, Higgs mass bounds, B physics and $g_{\mu}-2$ are compatible with Yukawa unification within 10$\%$ in the 4-2-2 model. 

In $R_{tb\tau}$ - $A_{0}/m_{0}$ plane we can see that almost all the range is allowed form sparticle and Higgs mass bounds. On the other hand the B physics and $g_{\mu}-2$ constraints do not allow $A_{0}/m_{0}>-1$ for $R_{tb\tau}\lesssim 1.3$ \footnote{As a matter of fact $R_{tb\tau}\lesssim1.3$ requires $A_{0}/m_{0}<-1$. However, for higher value of $R_{tb\tau}$, $A_{0}/m_{0}$ is essentially unbounded.}.

In $R_{tb\tau}$ - $m_{H_{d}}/m_{H_{u}}$ plane we can see that for $R_{tb\tau }$ $\sim$ $1$ the ratio $m_{H_{d}}/m_{H_{u}}$ should be $\sim 1.1$ due to the B physics and $g_{\mu}-2$ constraints. It can also be seen that if we restrict ourselves to only 5 $\%$ or better Yukawa unification $R_{tb\tau}\lesssim 1.05$ then the allowed range is $1.05$ $\lesssim$ $m_{H_{d}}/ m_{H_{u}}$ $\lesssim$ $1.3$ and is compatible with B physics and $g_{\mu}-2$ constraints.   

\begin{figure}[H]
	\captionsetup{width=0.90\textwidth}
	\begin{subfigure}{.5\textwidth}
		\includegraphics[width=8cm,height=6cm]{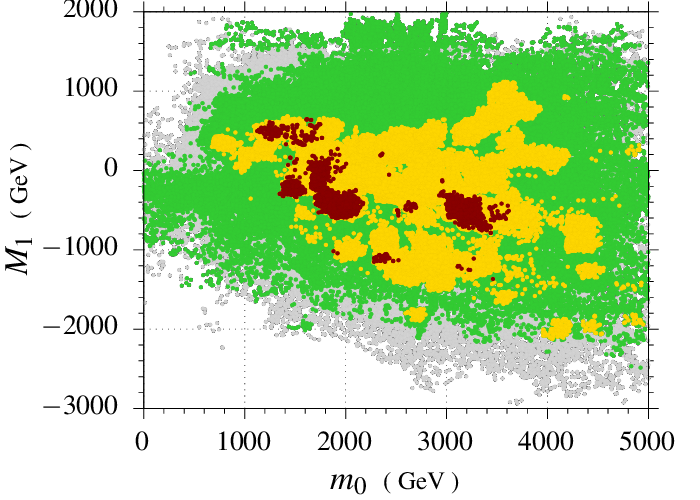}
		\label{fig:sfig21}
	\end{subfigure}
	\begin{subfigure}{.5\textwidth}
		\includegraphics[width=8cm,height=6cm]{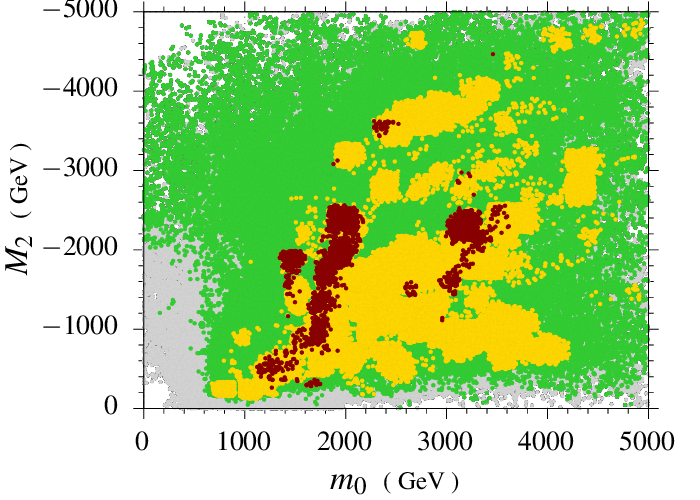}
		\label{sfig22}
	\end{subfigure}
	\begin{subfigure}{.5\textwidth}
		\includegraphics[width=8cm,height=6cm]{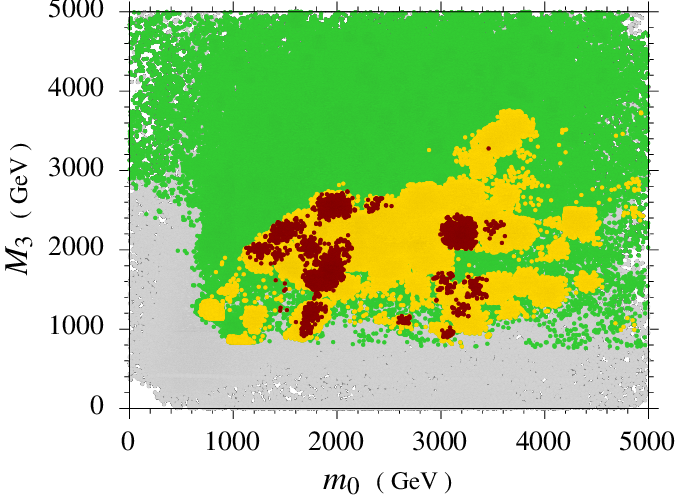}
		\label{fig:sfig23}
	\end{subfigure}
	\begin{subfigure}{.5\textwidth}
		\includegraphics[width=8cm,height=6cm]{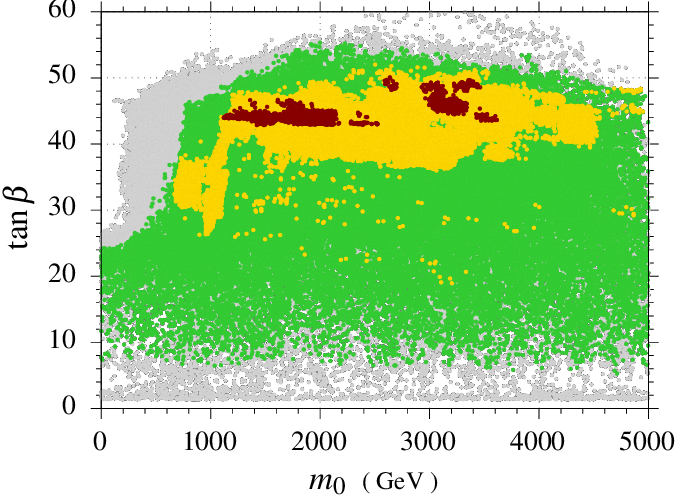}
		\label{fig:sfig24}
	\end{subfigure}
	\caption{Red points are a subset of yellow points and satisfy the $t$-$b$-$\tau$ Yukawa unification within  $10 \% $ i.e. $R_{tb\tau}<1.1$. Rest of the color coding is same as in Fig. (\ref{fig:fig1})}
	\label{fig:fig2}
\end{figure}

In Fig. (\ref{fig:fig2}), we present our results in $m_{0}$-$M_{1}$, $m_{0}$-$M_{2}$, $m_{0}$-$M_{3}$ and $m_{0}$-$\tan \beta$ planes. The red color points correspond to a subset of yellow color points and additionally satisfy Yukawa unification within 10 $\%$ range. In the $m_{0}$-$M_{1}$ plane, we can see that for lower sfermion masses, $m_{0} \lesssim 300$ GeV, the allowed mass range for $M_{1}$ is $-1400$ GeV $\lesssim$ $M_{1}$ $\lesssim$ $900$ GeV. This range is further shrunk by the sparticle and Higgs mass ranges to $-1000$ GeV $\lesssim$ $M_{1}$ $\lesssim$ $250$ GeV. B physics and $g_{\mu}-2$ constraints do not favor this range. It can also be observed that B physics and $g_{\mu}-2$ constraints put a lower bound of $\sim 600$ GeV on $m_{0}$. The gaugino mass parameter bounds from B physics and $g_{\mu}-2$ constraints are $-2.1$ TeV $\lesssim$ $M_{1}$ $\lesssim$ $1.2$ TeV. In general, we can see two opposite trends, the sparticle and Higgs mass bounds seem to prefer the positive $M_{1}$ while B physics and $g_{\mu}-2$ prefer negative $M_{1}$. This comes simply from the fact that the SUSY contribution to $g_\mu-2$ requires $M_2 <0$ for $\mu<0$, and so, pushes $M_1$ to negative values. 

In the $m_{0}$ - $M_{2}$ plane we can see that for $m_{0} \lesssim 400$ GeV, $M_{2}\geq -1000$ GeV is not allowed by sparticle and Higgs mass bounds. For B physics and $g_{\mu}-2$ constraints, the allowed range of $M_{2}$ increases for heavier $m_{0}$. The missing region on the lower left corner of the plot is due to the Higgs mass constraint given in Eq.\eqref{higgs}.

In the $m_{0}$-$M_{3}$ plane, we can see that the B physics constraints put an upper limit of $3$ TeV on $M_{3}$ for $m_{0}$ $\lesssim $ $2$ TeV. The hard cut at $M_{3} \sim 900$ GeV is essentially due to the gluino mass bounds while the missing region over the left half is due to the Higgs mass bound. B physics and $g_{\mu}-2$ constraints provide an upper limit of $\sim 4$ TeV on $M_{3}$ for the $4$-$2$-$2$ model.
In $m_{0}$-$\tan\beta$ plane, we can see that the sparticle and (primarily) the Higgs mass bounds do not allow $\tan \beta$ $\lesssim$ $6$. The sharp green cut in the upper left-half is due to Higgs mass constraints given in Eq.\eqref{higgs}. B physics and $g_{\mu}-2$ constraints do not allow $\tan\beta \lesssim 17$ for the $4$-$2$-$2$ model and these constraints also provide an upper limit of $\tan\beta \sim 51$. 

\begin{figure}[H]
	\captionsetup{width=0.90\textwidth}
	\begin{subfigure}{.5\textwidth}
		\centering
		\includegraphics[width=8cm,height=6cm]{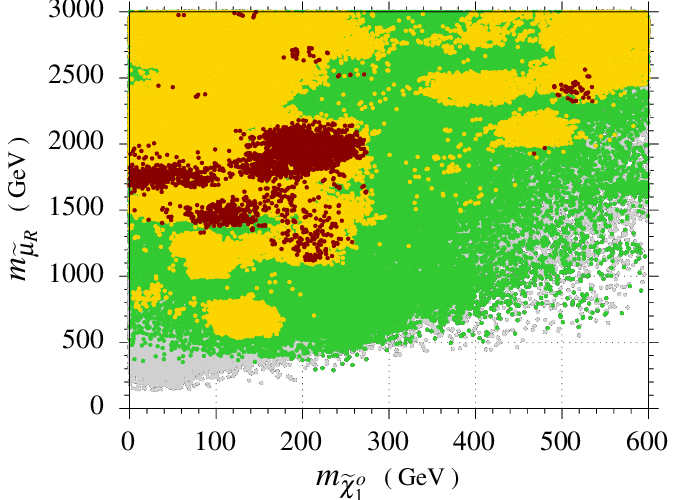}
		\label{fig:sfig31}
	\end{subfigure}
	\begin{subfigure}{.5\textwidth}
		\centering
		\includegraphics[width=8cm,height=6cm]{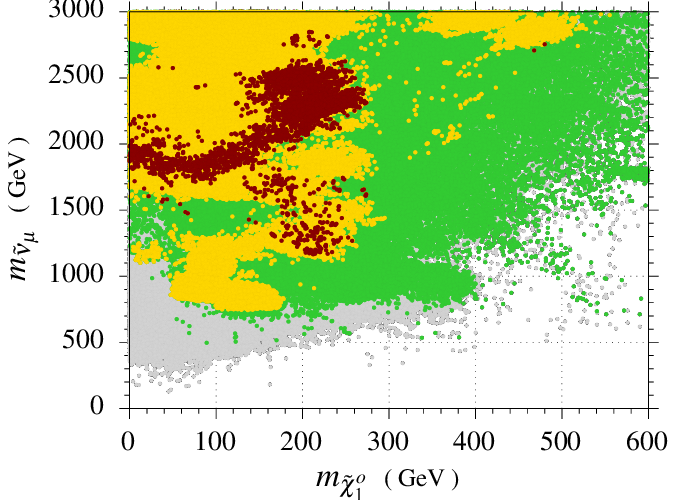}
		\label{fig:sfig32}
	\end{subfigure}
	\begin{subfigure}{.5\textwidth}
		\centering
		\includegraphics[width=8cm,height=6cm]{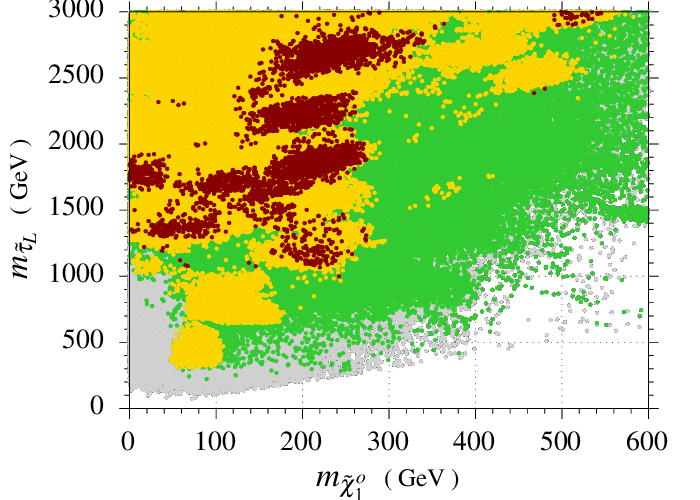}
		\label{fig:sfig33}
	\end{subfigure}
	\begin{subfigure}{.5\textwidth}
		\centering
		\includegraphics[width=8cm,height=6cm]{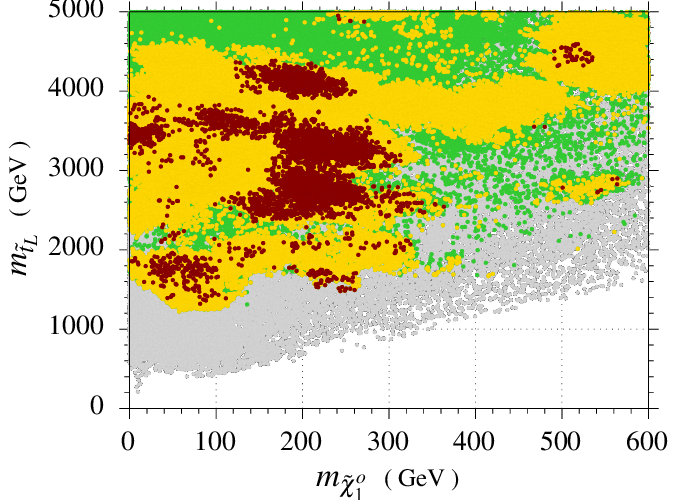}
		\label{fig:sfig34}
	\end{subfigure}
	\caption{Plots in the  $m_{\tilde{\mu}_{R}}$ - $m_{\tilde{\chi}^{1}_{0}}$ and $m_{\tilde{\nu}_{\mu}}$ - $m_{\tilde{\chi}^{1}_{0}}$, $m_{\tilde{\tau}_{L}}$ - $m_{\tilde{\chi}^{1}_{0}}$ and $m_{\tilde{t}_{L}}$ - $m_{\tilde{\chi}^{1}_{0}}$ planes, See Fig. (\ref{fig:fig2}) for 
		details of the colors used.}
	\label{fig:fig3}
\end{figure}

In Fig. (\ref{fig:fig3}) we discuss representative sparticle spectroscopy. We can see that all constraints are consistent with $10 \%$ Yukawa unification. Its further  clear from Fig. (\ref{fig:fig3}) that we can not have $\tilde{\mu}_{R}$,  $\tilde{\nu}_{\mu}$, $\tilde{\tau}_{L}$, $\tilde{t}_{L}$ co-annihilation with $\tilde{\chi}^{1}_{0}$ in the 4-2-2 model. $\tilde{\mu}_{R}$ is the supersymmetric partner of right-handed muon, $\tilde{\nu}_{\mu}$ is the supersymmetric partner of muon neutrino, $\tilde{\tau_{L}}$ is the supersymmetric partner of left handed $\tau$ - lepton and $\tilde{t}_{L}$ is the supersymmetric partner of left handed top quark. 

In $m_{\tilde{\mu}_{R}}$ - $m_{\tilde{\chi}^{1}_{0}}$ plot we can see that B physics and $g_{\mu}-2$ anomaly constraints put a lower limit of $\sim 500$ GeV on $m_{\tilde{\mu}_{R}}$ whereas 10 $\%$ or better Yukawa unification occurs for $m_{\tilde{\mu}_{R}}$ $\geq$ $1100$ GeV.

In $m_{\tilde{\nu}_{\mu}}$ - $m_{\tilde{\chi}^{1}_{0}}$ plane we can infer that $g_{\mu}-2$ and B physics constraints set a lower mass bound of $\sim 700$ GeV on mass of smuon-neutrino. Yukawa unification is possible for $m_{\tilde{\nu}_\mu}$~$\geq$~$1100$ GeV. In $m_{\tilde{\tau}_{L}}$ - $m_{\tilde{\chi}^{1}_{0}}$ plane we can see that the mass bounds sparticle and the Higgs (mainly the Higgs mass bound, given in Eq.\eqref{higgs}) provide a lower mass limit of $\sim700$ GeV on $m_{\tilde{\tau}_{L}}$ for $m_{\tilde{\chi}^{0}_{1}}$ $\lesssim$ $200$ GeV.  In this plot we can also see that the Yukawa unification is possible for $m_{\tilde{\tau}_{L}}$ $\geq$ $1000$ GeV. We can also see in this figure that B physics constraints impose a lower mass bound of $\sim 300$ GeV on $m_{\tilde{\tau}_{L}}$.
In $m_{\tilde{t}_{L}}$ - $m_{\tilde{\chi}^{1}_{0}}$ plane we can see that B physics and $g_{\mu}-2$ anomaly constraints put a lower limit of $\sim1200$ GeV on $m_{\tilde{t}_{L}}$. This lower limit is also compatible with Yukawa unification.

\begin{figure}[H]
	\captionsetup{width=0.90\textwidth}
	\begin{subfigure}{.5\textwidth}
		\includegraphics[width=8cm,height=6cm]{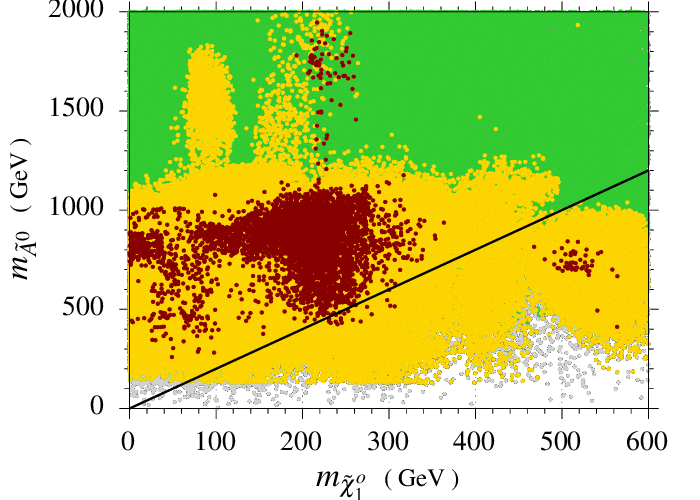}
		\label{fig:sfig41}
	\end{subfigure}
	\begin{subfigure}{.5\textwidth}
		\includegraphics[width=8cm,height=6cm]{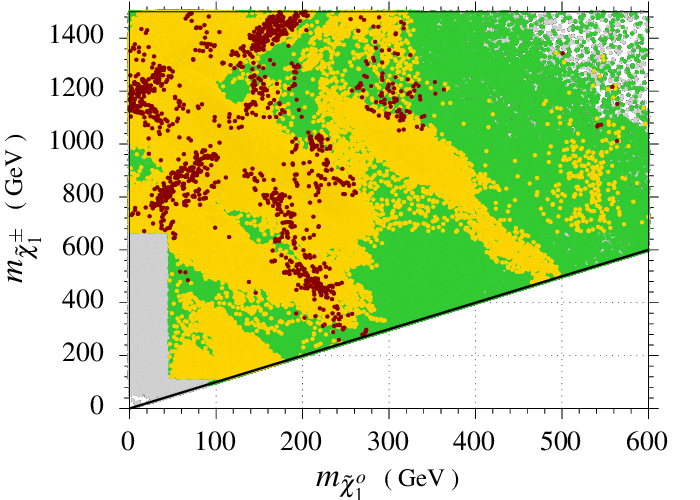}
		\label{sfig42}
	\end{subfigure}
	\caption{Plots in $m_{{A}^{0}}$ - $m_{\tilde{\chi}^{0}_{1}}$ and $m_{\tilde{\chi}^{\pm}_{1}}$ - $m_{\tilde{\chi}^{0}_{1}}$ planes. Slope line in $m_{\tilde{A}^{0}}$ - $m_{\tilde{\chi}^{0}_{1}}$ plot is for $m_{\tilde{A}^{0}}$ = $2m_{\tilde{\chi}^{0}_{1}}$. See Fig. (\ref{fig:fig2}) for 
	details of the colors used.}
	\label{fig:fig4}
\end{figure}

In Fig. (\ref{fig:fig4}) we show plots in the $m_{A^{0}}$ - $m_{\chi^{0}_{1}}$, $m_{\chi^{\pm}_{1}}$ - $m_{\chi^{0}_{1}}$ planes, where $m_{A^{0}}$ is mass of neutral pseudo scalar Higgs, and $m_{\chi^{\pm}_1}$ is the lightest chargino, a combination of winos and higgsinos (the supersymmetric partners of $W^{\pm}$ bosons and Higgs boson). 

In $m_{A^{0}}$ - $m_{\chi^{0}_{1}}$ plane we find that the $A^{0}$ resonance region is compatible with B physics and $g_{\mu}-2$ constraints. It is also compatible with $10$ $\%$ or better Yukawa unification. 

In $m_{\chi^{\pm}_{1}}$ - $m_{\chi^{0}_{1}}$ plane we can see the chargino-neutralino resonance region compatible with all constraints mentioned in the previous section. The hard cuts are due to neutralino and chargino bounds given in Eq.\eqref{gluino}.
In Table \ref{t1} we present the dark matter matter relic density calculation for two example points, first one is for $A^{0}$ resonance and second one is for $\tilde{\chi}^{\pm}_{1}$ co-annihilation scenario, which satisfy all theoretical and experimental constraints given in previous section.
\begin{table}[H]
	\captionsetup{width=0.95\textwidth}
	\footnotesize
	
	\centering
	\begin{tabular}{|ccccccc|ccccc|c|}
		\hline
		{$m_{0}$} & {$A_{0}$} & {$M_{2}$} & {$M_{3}$} &
		{$m_{H_{d}}$} & {$m_{H_{u}}$} & {$\tan \beta $} & {$m_{{A}^{0}}$} & {$|m_{\tilde{\chi}^{0}_{1}}|$}& {$|m_{\tilde{\chi}^{\pm}_{1}}|$}& {$R_{tb\tau}$} & {$m_{\tilde{g}}$} & {$\Omega h^{2}$} \\
		\hline
		\hline
		{3310} & {-8160} & {-2450} & {2225} &
		{4687} & {4127} & {45.8} & {533} & {263} & {2074} & {1.09} & {4776} & {0.1} \\
		{1270} & {-3200} & {-263} & {1830} &
		{1898} & {1896}  & {45.4}& {866}  & {238} & {259} &{1.06} & {3904} & {0.1}\\
		\hline
	\end{tabular}
	\caption{$A^{0}$ resonance($1^{st}$ row) and $\tilde{\chi}^{\pm}_{1}$ co-annihilation ($2^{nd}$ row). All masses are in GeV.}
	\label{t1}
\end{table}
 We have also calculated the spin independent and spin dependent cross sections for the two points and th corresponding values are given in Table~\ref{t2} below.
\begin{table}[H]
	\captionsetup{width=0.95\textwidth}
	\footnotesize
	
	\centering
	\begin{tabular}{|ccccccc|ccc|}
		\hline
		{$m_{0}$} & {$A_{0}$} & {$M_{2}$} & {$M_{3}$} &
		{$m_{H_{d}}$} & {$m_{H_{u}}$} & {$\tan \beta $} &{$\sigma_{SI}$($cm^{2}$)}&{$\sigma_{SD}$($cm^{2}$)}& {$\langle \sigma_{A} v \rangle$ ($cm^3/s$)}   \\
		\hline
		\hline
		{3310} & {-8160} & {-2450} & {2225} &
		{4687} & {4127} & {45.8}&{4.9 $\times 10^{-47}$}& $2.0 \times 10^{-45}$& {$1.6 \times 10^{-26}$}\\
		{1270} & {-3200} & {-263} & {1830} &
		{1898} & {1896}  & {45.4}&{$5.9 \times 10^{-48}$}& {$4.5 \times 10^{-45}$}& {$9.4 \times 10^{-29}$}\\
		\hline
	\end{tabular}
	\caption{Spin independent cross section, spin dependent cross section and velocity averaged product of self annihilation cross section and velocity of dark matter for the two points which have correct relic abundance.}
	\label{t2}
\end{table}
The XENON1t experiment \cite{Aprile:2018dbl} provides the latest value for the upper bound of $\sigma_{SI}$ to be of the order of $\sim 10^{-47}$ $cm^{2}$ for a dark matter particle of mass 30 GeV and this bound shifts up to $\sim 10^{-46}$ $cm^{2}$ for a dark matter particle of mass $\sim 200-300$ GeV which is the case for our calculated points given in Table \ref{t1}. The spin dependent cross section is experimentally measured by the Fermi-LAT collaboration \cite{Ajello:2011dq} and they have provided an upper bound of the order of $10^{-44}$ $cm^{2}$ for a dark matter particle of $\sim 200-300$ GeV which is one order higher than the order of the spin-dependent cross section calculated for our two example points given in above tables. The last column in Table \ref{t2} gives the velocity averaged product of the dark matter self annihilation cross section and the relative velocity of dark matter particles $\langle \sigma_{A} v \rangle$ which is being measured experimentally by the Fermi-LAT and ICECUBE experiments. The latest exclusion limit \cite{Ackermann:2015tah, Aartsen:2017ulx} on $\langle \sigma_{A} v \rangle$ is $\sim 10^{-25}$ $cm^3/s$ for a dark matter particle with mass $\sim 200$ GeV. Our sample data points are within the allowed region and may be tested in ongoing and future dark matter searches.  
\begin{figure}[H]
	\captionsetup{width=0.90\textwidth}
	\begin{subfigure}{.5\textwidth}
		\includegraphics[width=8cm,height=6cm]{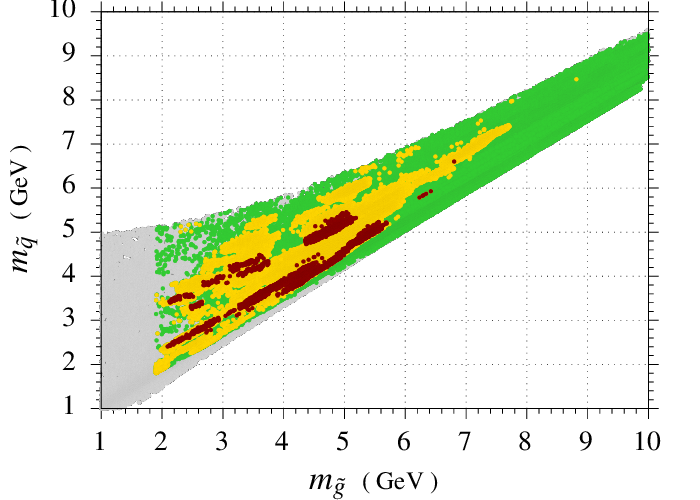}
		\label{fig:sfig51}
	\end{subfigure}
	\begin{subfigure}{.5\textwidth}
		\includegraphics[width=8cm,height=6cm]{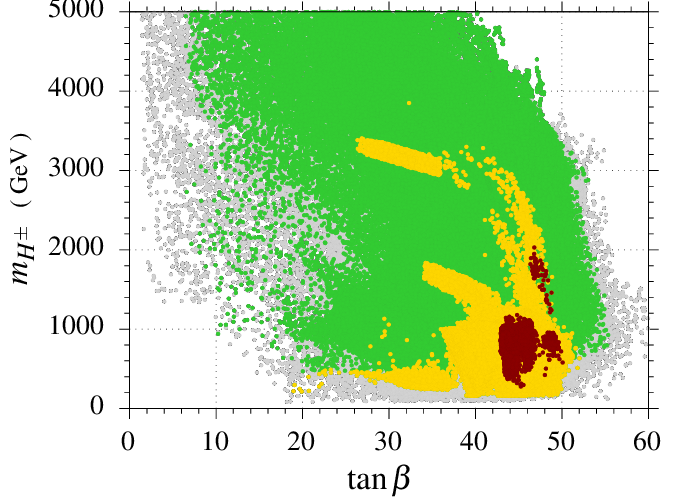}
		\label{sfig52}
	\end{subfigure}
	\caption{Plots in $m_{\tilde{g}}$ - $m_{\tilde{q}}$ and $m_{H^{\pm}}$ - $\tan \beta$ planes. See Fig. (\ref{fig:fig2}) for 
		details of the colors used.}
	\label{fig:fig5}
\end{figure}
In Fig. (\ref{fig:fig5}) we show our plots in $m_{\tilde{g}}$ - $m_{\tilde{q}}$ and $m_{H^{\pm}}$ - $\tan \beta$ planes where $m_{H^{\pm}}$ is the mass of charged Higgs particles. 

In $m_{\tilde{g}}$ - $m_{\tilde{q}}$ plot we can see that heavy squarks and heavy gluino masses are compatible with B physics constraints as expected. We can also get 10 $\%$ or better Yukawa unification with heavy gluino and squarks.

In $m_{H^{\pm}}$ - $\tan \beta$ plot we can see that B physics and $g_{\mu}-2$ constraints prefer low mass of $H^{\pm}$ particles. It is due to the fact that the MSSM corrections to the B physics observables inversely depend on the charged Higgs particles masses. 

\section{Conclusion}
\label{conclusion}
We have seen that in the supersymmetric 4-2-2 model, constraints from current collider bounds on supersymmetric particles, from the Higgs mass bound,  from the rare decays of B physics and from the muon $g_{\mu}-2$ anomaly are satisfied along with $10\%$ or better third family Yukawa unification.
These models also lead to scenarios with the correct dark matter relic density in addition to all latest experimental bounds on spin-independent and spin dependent scattering cross sections for 
neutralino as the dark matter candidate. 

We have seen that for 4-2-2 model,  $m_{0}$ should be greater than $600$ GeV. This limit is provided by all constraints mentioned in section\ref{scanProc} with 10 $\%$ or better Yukawa unification. It has also been shown that 10 $\%$ or better Yukawa unification is possible for $42 \lesssim \tan \beta \lesssim 50$ and for this range, all other constraints, experimental and theoretical, are also satisfied. The lower mass bounds on different sparticle masses are also been given. Right handed smuon and the smuon neutrino has mass limit of 500 GeV and 700 GeV, respectively. For 10 $\%$ or better Yukawa unification the lower mass bound on right handed smuon and the smuon neutrino is 1100 GeV. All constraints with 10 $\%$ or better Yukawa unification put a lower mass limit of 700 GeV and 1200 GeV on $m_{\tilde{\tau}_{L}}$ and $m_{\tilde{t}_{L}}$ respectively.

All experimental constraints provided by sparticle mass bounds, rare decays of B meson, $g_{\mu}-2$ anomaly and dark matter scattering cross section bounds are being satisfied with a TeV mass spectrum of sparticles which gives a hint for the detection of supersymmetric particle in the future collider experiments.  
\section*{Acknowledgment}
We acknowledge the use of super-computing facility at National Centre for Physics (NCP), Shahdra valley road, Islamabad and at Research Centre for Modeling and Simulations(RCMS) at National University of Sciences and Technology (NUST), H-12, Islamabad for calculation of results presented in this paper and thank the same.

\end{document}